\documentstyle[prd,floats,aps,epsf,twocolumn]{revtex}
\input epsf

\newcounter{itemlistc}
\newcounter{romanlistc}
\newcounter{alphlistc}
\newcounter{arabiclistc}

\newcommand {\abs}[1]{\mid \!\! #1 \!\! \mid}
\def \cal {\mathcal}
\def \met {{\,/\!\!\!\!E_{T}}}

\def \begineq {\begin{equation}}
\def \endeq {\end{equation}}
\def \Nmin {{N_{\rm min}}} 
\def \scriptR {\mbox{${\cal R}$}}
\def \scriptP {\mbox{${\cal P}$}}
\def \gothicP {\tilde{\scriptP}}

\def \Sleuth   {{Sleuth}}
\def \Sherlock {\Sleuth}
\def \hse {{\small{hse}}}
\def \hser {{\small{hser}}}
\def \ellgamma {\mbox{($\ell/\gamma$)}}
\def \jj {\,2j}
\def \jjj {\,3j}
\def \jjjj {\,4j}
\def \jjjjj {\,5j}
\def \jjjjjj {\,6j}

\newcommand{\guess}[1]
{\underline{#1}}

\newcommand{\dofig}[4]
{\begin{figure}[htbp]
\begin{center}
\leavevmode
\hbox{%
\epsfxsize=#2
\epsfbox{#1}}
\caption{#3}
\label{#4}
\end{center}
\end{figure}}

\def \scriptPemumet {$0.14$}
\def \scriptPemumetsigma {$+1.08$}
\def \scriptPemumetj {$0.45$}
\def \scriptPemumetjsigma {$+0.13$}
\def \scriptPemumetjj {$0.31$}
\def \scriptPemumetjjsigma {$+0.50$}
\def \scriptPemumetjjj {$0.71$}
\def \scriptPemumetjjjsigma {$-0.55$}

\def \scriptPeejj {$0.72$}
\def \scriptPeejjsigma {$-0.58$}
\def \scriptPeejjj {$0.61$}
\def \scriptPeejjjsigma {$-0.28$}
\def \scriptPeejjjj {$0.04$}
\def \scriptPeejjjjsigma {$+1.75$}

\def \scriptPeemetjj {$0.68$}
\def \scriptPeemetjjsigma {$-0.47$}
\def \scriptPeemetjjj {$0.36$}
\def \scriptPeemetjjjsigma {$+0.36$}
\def \scriptPeemetjjjj {$0.06$}
\def \scriptPeemetjjjjsigma {$+1.55$}

\def \scriptPmumujj {$0.08$}
\def \scriptPmumujjsigma {$+1.41$}

\def \scriptPZjj {$0.52$}
\def \scriptPZjjsigma {$-0.05$}
\def \scriptPZjjj {$0.71$}
\def \scriptPZjjjsigma {$-0.55$}
\def \scriptPZjjjj {$0.83$}
\def \scriptPZjjjjsigma {$-0.95$}

\def \scriptPemetjj {$0.76$}
\def \scriptPemetjjsigma {$-0.71$}
\def \scriptPemetjjj {$0.17$}
\def \scriptPemetjjjsigma {$+0.95$}
\def \scriptPemetjjjj {$0.13$}
\def \scriptPemetjjjjsigma {$+1.13$}

\def \scriptPWjj {$0.29$}
\def \scriptPWjjsigma {$+0.55$}
\def \scriptPWjjj {$0.23$}
\def \scriptPWjjjsigma {$+0.74$}
\def \scriptPWjjjj {$0.53$}
\def \scriptPWjjjjsigma {$-0.08$}
\def \scriptPWjjjjj {$0.81$}
\def \scriptPWjjjjjsigma {$-0.88$}
\def \scriptPWjjjjjj {$0.22$}
\def \scriptPWjjjjjjsigma {$+0.77$}

\def \scriptPeeg {$0.88$}
\def \scriptPeegsigma {$-1.17$}

\def \scriptPeegmet {$0.23$}
\def \scriptPeegmetsigma {$+0.74$}
\def \scriptPZg {$0.84$}
\def \scriptPZgsigma {$-0.99$}
\def \scriptPZgj {$0.63$}
\def \scriptPZgjsigma {$-0.33$}
\def \scriptPeee {$0.89$}
\def \scriptPeeesigma {$-1.23$}

\def \scriptPegg {$0.66$}
\def \scriptPeggsigma {$-0.41$}
\def \scriptPeggj {$0.21$}
\def \scriptPeggjsigma {$+0.81$}
\def \scriptPeggjj {$0.30$}
\def \scriptPeggjjsigma {$+0.52$}
\def \scriptPWgg {$0.18$}
\def \scriptPWggsigma {$+0.92$}
\def \scriptPggg {$0.41$}
\def \scriptPgggsigma {$+0.23$}

\def \twiddleScriptPvalue {$0.89$}
\def \twiddleScriptPvalueSigma {$-1.23$}

\begin{document}

\onecolumn
\title{A Quasi-Model-Independent Search for New High $p_T$ Physics at D\O}

%
\author{                                                                      
B.~Abbott,$^{56}$                                                             
A.~Abdesselam,$^{11}$                                                         
M.~Abolins,$^{49}$                                                            
V.~Abramov,$^{24}$                                                            
B.S.~Acharya,$^{16}$                                                          
D.L.~Adams,$^{58}$                                                            
M.~Adams,$^{36}$                                                              
G.A.~Alves,$^{2}$                                                             
N.~Amos,$^{48}$                                                               
E.W.~Anderson,$^{41}$
% ...
\\                                                                            
\vskip 0.30cm                                                                 
\centerline{(D\O\ Collaboration)}                                             
\vskip 0.30cm                                                                 
}                                                                             
\address{                                                                     
\centerline{$^{1}$Universidad de Buenos Aires, Buenos Aires, Argentina}       
\centerline{$^{2}$LAFEX, Centro Brasileiro de Pesquisas F{\'\i}sicas,         
                  Rio de Janeiro, Brazil}                                     
\centerline{$^{3}$Universidade do Estado do Rio de Janeiro,                   
                  Rio de Janeiro, Brazil}                                     
\centerline{$^{4}$Institute of High Energy Physics, Beijing,                  
                  People's Republic of China}                                 
\centerline{$^{5}$Universidad de los Andes, Bogot\'{a}, Colombia}             
\centerline{$^{6}$Charles University, Prague, Czech Republic}                 
\centerline{$^{7}$Institute of Physics, Academy of Sciences, Prague,          
                  Czech Republic}                                             
\centerline{$^{8}$Universidad San Francisco de Quito, Quito, Ecuador}         
\centerline{$^{9}$Institut des Sciences Nucl\'eaires, IN2P3-CNRS,             
                  Universite de Grenoble 1, Grenoble, France}                 
\centerline{$^{10}$CPPM, IN2P3-CNRS, Universit\'e de la M\'editerran\'ee,     
                  Marseille, France}                                          
\centerline{$^{11}$Laboratoire de l'Acc\'el\'erateur Lin\'eaire,              
                  IN2P3-CNRS, Orsay, France}                                  
\centerline{$^{12}$LPNHE, Universit\'es Paris VI and VII, IN2P3-CNRS,         
                  Paris, France}                                              
\centerline{$^{13}$DAPNIA/Service de Physique des Particules, CEA, Saclay,    
                  France}                                                     
\centerline{$^{14}$Panjab University, Chandigarh, India}                      
\centerline{$^{15}$Delhi University, Delhi, India}                            
\centerline{$^{16}$Tata Institute of Fundamental Research, Mumbai, India}     
\centerline{$^{17}$Seoul National University, Seoul, Korea}                   
\centerline{$^{18}$CINVESTAV, Mexico City, Mexico}                            
\centerline{$^{19}$FOM-Institute NIKHEF and University of                     
                  Amsterdam/NIKHEF, Amsterdam, The Netherlands}               
\centerline{$^{20}$University of Nijmegen/NIKHEF, Nijmegen, The               
                  Netherlands}                                                
\centerline{$^{21}$Institute of Nuclear Physics, Krak\'ow, Poland}            
\centerline{$^{22}$Institute for Theoretical and Experimental Physics,        
                   Moscow, Russia}                                            
\centerline{$^{23}$Moscow State University, Moscow, Russia}                   
\centerline{$^{24}$Institute for High Energy Physics, Protvino, Russia}       
\centerline{$^{25}$Lancaster University, Lancaster, United Kingdom}           
\centerline{$^{26}$Imperial College, London, United Kingdom}                  
\centerline{$^{27}$University of Arizona, Tucson, Arizona 85721}              
\centerline{$^{28}$Lawrence Berkeley National Laboratory and University of    
                  California, Berkeley, California 94720}                     
\centerline{$^{29}$University of California, Davis, California 95616}         
\centerline{$^{30}$California State University, Fresno, California 93740}     
\centerline{$^{31}$University of California, Irvine, California 92697}        
\centerline{$^{32}$University of California, Riverside, California 92521}     
\centerline{$^{33}$Florida State University, Tallahassee, Florida 32306}      
\centerline{$^{34}$University of Hawaii, Honolulu, Hawaii 96822}              
\centerline{$^{35}$Fermi National Accelerator Laboratory, Batavia,            
                   Illinois 60510}                                            
\centerline{$^{36}$University of Illinois at Chicago, Chicago,                
                   Illinois 60607}                                            
\centerline{$^{37}$Northern Illinois University, DeKalb, Illinois 60115}      
\centerline{$^{38}$Northwestern University, Evanston, Illinois 60208}         
\centerline{$^{39}$Indiana University, Bloomington, Indiana 47405}            
\centerline{$^{40}$University of Notre Dame, Notre Dame, Indiana 46556}       
\centerline{$^{41}$Iowa State University, Ames, Iowa 50011}                   
\centerline{$^{42}$University of Kansas, Lawrence, Kansas 66045}              
\centerline{$^{43}$Kansas State University, Manhattan, Kansas 66506}          
\centerline{$^{44}$Louisiana Tech University, Ruston, Louisiana 71272}        
\centerline{$^{45}$University of Maryland, College Park, Maryland 20742}      
\centerline{$^{46}$Boston University, Boston, Massachusetts 02215}            
\centerline{$^{47}$Northeastern University, Boston, Massachusetts 02115}      
\centerline{$^{48}$University of Michigan, Ann Arbor, Michigan 48109}         
\centerline{$^{49}$Michigan State University, East Lansing, Michigan 48824}   
\centerline{$^{50}$University of Nebraska, Lincoln, Nebraska 68588}           
\centerline{$^{51}$Columbia University, New York, New York 10027}             
\centerline{$^{52}$University of Rochester, Rochester, New York 14627}        
\centerline{$^{53}$State University of New York, Stony Brook,                 
                   New York 11794}                                            
\centerline{$^{54}$Brookhaven National Laboratory, Upton, New York 11973}     
\centerline{$^{55}$Langston University, Langston, Oklahoma 73050}             
\centerline{$^{56}$University of Oklahoma, Norman, Oklahoma 73019}            
\centerline{$^{57}$Brown University, Providence, Rhode Island 02912}          
\centerline{$^{58}$University of Texas, Arlington, Texas 76019}               
\centerline{$^{59}$Texas A\&M University, College Station, Texas 77843}       
\centerline{$^{60}$Rice University, Houston, Texas 77005}                     
\centerline{$^{61}$University of Virginia, Charlottesville, Virginia 22901}   
\centerline{$^{62}$University of Washington, Seattle, Washington 98195}       
}                                                                             

\maketitle
\vskip 10pt

{\samepage
{\bf
\begin{center}
Abstract
\end{center}
}
\begin{center}
\begin{minipage}{.8\textwidth}
{\small 

We apply a quasi-model-independent strategy (``\Sherlock'') to search for new high $p_T$ physics in $\approx 100$ pb$^{-1}$ of $p\bar{p}$ collisions at $\sqrt{s} = 1.8$~TeV collected by the D\O\ experiment during 1992--1996 at the Fermilab Tevatron.  We systematically analyze many exclusive final states and demonstrate sensitivity to a variety of models predicting new phenomena at the electroweak scale.  No evidence of new high $p_T$ physics is observed.

}
\end{minipage}
\end{center}
}

\vskip 2.0in

\twocolumn

\def \ellgamma {{(e/\mu/\gamma)}}


It is generally recognized that the standard model, an extremely successful description of the fundamental particles and their interactions, must be incomplete.  Unfortunately, the possibilities beyond the current paradigm are sufficiently broad that the first hint could appear in any of many different guises.  This suggests the importance of performing searches that are as model-independent as possible.  In this Letter we describe a search for new physics beyond the standard model, assuming nothing about the expected characteristics of the new processes other than that they will produce an excess of events at high transverse momentum ($p_T$).  An explicit prescription (``\Sleuth'')~\cite{SherlockPRD1,KnutesonThesis} is applied to many exclusive final states~\cite{SherlockPRD1,KnutesonThesis,SherlockPRD2} in a data sample corresponding to approximately $100$ pb$^{-1}$ of $p\bar{p}$ collisions collected by the D\O\ detector~\cite{D0Detector} during 1992--1996 (Run I) at the Fermilab Tevatron.


The data are partitioned into exclusive final states using standard criteria that identify isolated and energetic electrons ($e$), muons ($\mu$), and photons ($\gamma$), as well as jets ($j$), missing transverse energy ($\met$), and the presence of $W$ and $Z$ bosons~\cite{SherlockPRD1}.  For each exclusive final state, we consider a small set of variables given in Table~\ref{tbl:VariableRules}.  The notation $\sum'{p_T^j}$ is shorthand for $p_T^{j_1}$ if the final state contains only one jet, and $\sum_{i=2}^n{p_T^{j_i}}$ if the final state contains $n \geq 2$ jets, unless the final state contains only $n \geq 3$ jets and no other objects, in which case $\sum_{i=3}^n{p_T^{j_i}}$ is used.  Leptons and $\met$ from reconstructed $W$ or $Z$ bosons are not considered separately in the left-hand column.  Because the muon momentum resolution in Run I was modest, we define $\sum{p_T^\ell} = \sum{p_T^e}$ for events with one or more electrons and one or more muons, and we determine $\met$ from the transverse energy summed in the calorimeter, which includes the $p_T$ of electrons, but only a negligible fraction of the $p_T$ of muons.  When there are exactly two objects in an event (e.g., one $Z$ boson and one jet), their $p_T$ values are expected to be nearly equal, and we therefore use the average $p_T$ of the two objects.  When  there is only one object in an event (e.g., a single $W$ boson), we use no variables, and simply count the number of such events.

\begin{table}[htb]
\centering
\begin{tabular}{cc}
If the final state includes & then consider the variable \\ \hline
$\met$ & $\met$ \\ 
one or more charged leptons & $\sum{p_T^\ell}$ \\ 
one or more electroweak bosons & $\sum{p_T^{\gamma/W/Z}}$ \\ 
one or more jets & $\sum'{p_T^j}$ \\ 
\end{tabular}
\caption{A quasi-model-independently motivated list of interesting variables for any final state.  The set of variables to consider for any exclusive channel is the union of the variables in the second column for each row that pertains to that final state.}
\label{tbl:VariableRules}
\end{table}


The \Sleuth\ algorithm requires as input a data sample, a set of events modeling each background process $i$, and the number of background events $\hat{b}_i \pm \delta \hat{b}_i$ from each background process expected in the data sample.  From these we determine the region $\scriptR$ of greatest excess and quantify the degree $\scriptP$ to which that excess is interesting.  The algorithm itself, applied to each individual final state, consists of seven steps:  

\noindent (1) We construct a mapping from the $d$-dimensional variable space defined by Table~\ref{tbl:VariableRules} into the $d$-dimensional unit box (i.e., $[0,1]^d$) that flattens the total background distribution.  We use this to map the data into the unit box.  

\noindent (2) We define a ``region'' $R$ about a set of $N$ data points to be the volume within the unit box closer to one of the data points in the set than to any of the other data points in the sample.  The arrangement of data points themselves thus determines the regions.  A region containing $N$ data points is called an $N$-region.

\noindent (3) Each region contains an expected number of background events $\hat{b}_R$, numerically equal to the volume of the region $\times$ the total number of background events expected, and an associated systematic error $\delta\hat{b}_R$, which varies within the unit box according to the systematic errors assigned to each contribution to the background estimate.  We can therefore compute the probability $p_N^R$ that the background in the region fluctuates up to or beyond the observed number of events.  This probability is the first measure of the degree of interest of a particular region.

\noindent (4) The rigorous definition of regions reduces the number of candidate regions from infinity to $\approx 2^{N_{\rm data}}$.  Imposing explicit criteria on the regions that the algorithm is allowed to consider further reduces the number of candidate regions.  We apply geometric criteria that favor high values in at least one dimension of the unit box, and we limit the number of events in a region to fifty.  The number of remaining candidate regions is still sufficiently large that an exhaustive search is impractical, and a heuristic is employed to search for regions of excess.  In the course of this search, the $N$-region $\scriptR_N$ for which $p_N^R$ is minimum is determined for each $N$, and $p_N = \min_R{(p_N^R)}$ is noted.  

\noindent (5) In any reasonably-sized data set, there will always be regions in which the probability for $b_R$ to fluctuate up to or above the observed number of events is small.  We determine the fraction $P_N$ of {\em hypothetical similar experiments} (\hse's) in which $p_N$ found for the \hse\ is smaller than $p_N$ observed in the data by generating random events drawn from the background distribution and computing $p_N$ by following steps (1)--(4). 

\noindent (6) We define $P$ and $\Nmin$ by $P=P_\Nmin=\min_N{(P_N)}$, and identify $\scriptR = \scriptR_\Nmin$ as the most interesting region in this final state.  

\noindent (7) We use a second ensemble of \hse's to determine the fraction $\scriptP$ of \hse's in which $P$ found in the \hse\ is smaller than $P$ observed in the data.  The most important output of the algorithm is this single number $\scriptP$, which may loosely be said to be the ``fraction of hypothetical similar experiments in which you would see an excess as interesting as what you actually saw in the data.''  $\scriptP$ takes on values between zero and unity, with values close to zero indicating a possible hint of new physics.  The computation of $\scriptP$ rigorously takes into account the many regions that have been considered within this final state.

The smallest $\scriptP$ found in the many different final states considered ($\scriptP_{\rm min}$) determines $\gothicP$, the ``fraction of {\em hypothetical similar experimental runs} (\hser's) that would have produced an excess as interesting as actually observed in the data,'' where an \hser\ consists of one \hse\ for each final state considered.  $\gothicP$ is calculated by simulating an ensemble of hypothetical similar experimental runs, and noting the fraction of these \hser's in which the smallest $\scriptP$ found is smaller than the smallest $\scriptP$ observed in the data.  Like $\scriptP$, $\gothicP$ takes on values between zero and unity, and the potential presence of new high $p_T$ physics would be indicated by finding $\gothicP$ to be small.  The difference between $\gothicP$ and $\scriptP$ is that in computing $\gothicP$ we account for the many final states that have been considered.  The correspondence between $\scriptP_{\rm min}$ and $\gothicP$ for the final states considered here is shown in Fig.~\ref{fig:results_plot_prl}(a). 


D\O\ has previously analyzed several final states ($\jj$, $ee$, $e\met$, $W\gamma$, $W$, $Z$, $Zj$, and $Wj$)~\cite{PreviousDZeroSearches} in a manner similar to the strategy used here, but without the benefit of \Sleuth.  No evidence of physics beyond the standard model was observed.  The final states we describe in this Letter divide naturally into four sets:  those containing one electron and one muon ($e\mu X$); those containing a single lepton, missing transverse energy, and two or more jets ($W$+jets-like); those containing two same-flavor leptons and two or more jets ($Z$+jets-like); and those in which the sum of the number of electrons, muons, and photons is $\geq 3$ [$3\ellgamma X$].



The $e\mu X$ data correspond to 108$\pm$6 pb$^{-1}$ of integrated luminosity.  The data and basic selection criteria are identical to those used in the published $t\bar{t}$ cross section analysis for the dilepton channels~\cite{topCrossSection}, which include the selection of events containing one or more isolated electrons with $p_T^e>15$~GeV, and one or more isolated muons with $p_T^\mu>15$~GeV.  In this Letter all electrons (and photons) have $\abs{\eta_{\rm det}}<1.1$ or $1.5<\abs{\eta_{\rm det}}<2.5$, and muons have $\abs{\eta_{\rm det}}<1.7$, unless otherwise indicated~\cite{Pseudorapidity}.  The dominant backgrounds to the $e\mu X$ final states are from $Z/\gamma^*\rightarrow \tau\tau \rightarrow e\mu\nu\nu\nu\nu$, and processes that generate a true muon and a jet that is misidentified as an electron.  Smaller backgrounds include $WW$ and $t\bar{t}$ production.



The $W$+jets-like final states include events in both the electron and muon channels.  The $e\met \jj(nj)$ events~\cite{LeptoquarksToENu}, corresponding to $115 \pm 6$~pb$^{-1}$ of collider data, have one electron with $p_T^e > 20$~GeV, $\met>30$~GeV, and two or more jets with $p_T^j>20$~GeV and $\abs{\eta_{\rm det}}<2.5$.  The electron and missing transverse energy are combined into a $W$ boson if $30<M_T^{e\nu}<110$~GeV. 
The $\mu\met \jj(nj)$ data~\cite{LeptoquarksToMuNu} correspond to $94 \pm 5$~pb$^{-1}$ of integrated luminosity.  Events in the final sample must contain one muon with $p_T^\mu>25$~GeV and $\abs{\eta_{\rm det}}<0.95$, two or more jets with $p_T^j>15$~GeV and $\abs{\eta_{\rm det}}<2.0$ and with the most energetic jet within $\abs{\eta_{\rm det}}<1.5$, and $\met>30$~GeV.  Because an energetic muon's momentum is not well measured in the detector, we are unable to separate ``$W$-like'' events from ``non-$W$-like'' events using the transverse mass, as done above in the electron channel.  The muon and missing transverse energy are therefore always combined into a $W$ boson.  The $W(\rightarrow\mu\met) \jj(nj)$ final states are combined with the $W(\rightarrow e\met) \jj(nj)$ final states described above to form the $W \jj(nj)$ final states.  
The dominant background to both the $e\met \jj(nj)$ and $\mu\met \jj(nj)$ final states is from $W$ + jets production.  A few events from $t\bar{t}$ production and semileptonic decay are expected in the final states $W\jjj$ and $W\jjjj$.  


The $Z$+jets-like final states also include events in both the electron and muon channels.  The $ee\jj(nj)$ data~\cite{LeptoquarksToEE} correspond to an integrated luminosity of $123 \pm 7$~pb$^{-1}$.  Offline event selection requires two electrons with transverse momenta $p_T^e > 20$~GeV and two or more jets with $p_T^j>20$~GeV and $\abs{\eta_{\rm det}}<2.5$.  We use a likelihood method to help identify events with significant missing transverse energy~\cite{SherlockPRD2}.  An electron pair is combined into a $Z$ boson if $82<M_{ee}<100$~GeV, unless the event contains significant $\met$ or a third charged lepton.  
The $\mu\mu \jj(nj)$ data~\cite{LeptoquarksToMuMu} correspond to $94 \pm 5$~pb$^{-1}$ of integrated luminosity.  Events in the final sample contain two or more muons with $p_T^\mu>20$~GeV and at least one muon with $\abs{\eta_{\rm det}}<1.0$, and two or more jets with $p_T^j>20$~GeV and $\abs{\eta_{\rm det}}<2.5$. A $\mu\mu$ pair is combined into a $Z$ boson if the muon momenta can be varied within their resolutions such that $m_{\mu\mu}\approx M_Z$ and $\met\approx 0$.
The dominant background to both the $ee\jj(nj)$ and $\mu\mu \jj(nj)$ data is from Drell-Yan production, with $Z/\gamma^*\rightarrow (ee/\mu\mu)$.


Events in the $3\ellgamma X$ final states are analyzed using $123 \pm 7$~pb$^{-1}$ of integrated luminosity.  All objects (electrons, photons, muons, and jets) are required to be isolated, to have $p_T\geq 15$~GeV, and to be within the fiducial volume of the detector.  Jets are required to have $\abs{\eta}<2.5$.  $\met$ is identified if its magnitude is larger than 15~GeV.  The dominant backgrounds to many of these final states include $Z\gamma$ and $WZ$ production.


\def \eejjOOdy {$20 \pm 4$}
\def \eejjOOqcd {$12.2 \pm 1.8$}
\def \eejjOOtotal {$32 \pm 4$}
\def \eejjOOdata {$32$}
\def \eejjjOOdy {$2.6 \pm 0.6$}
\def \eejjjOOqcd {$1.85 \pm 0.28$}
\def \eejjjOOtotal {$4.5 \pm 0.6$}
\def \eejjjOOdata {$4$}
\def \eejjjjOOdy {$0.40 \pm 0.20$}
\def \eejjjjOOqcd {$0.24 \pm 0.04$}
\def \eejjjjOOtotal {$0.64 \pm 0.20$}
\def \eejjjjOOdata {$3$}
\def \eejjjjjOOdy {$0.030 \pm 0.015$}
\def \eejjjjjOOqcd {$0.026 \pm 0.005$}
\def \eejjjjjOOtotal {$0.056 \pm 0.016$}
\def \eejjjjjOOdata {$0$}
\def \eejjjjjjOOdy {$0.0060 \pm 0.0030$}
\def \eejjjjjjOOqcd {$0.018 \pm 0.003$}
\def \eejjjjjjOOtotal {$0.024 \pm 0.005$}
\def \eejjjjjjOOdata {$0$}
\def \eejjjjjjjOOdy {$0.0010 \pm 0.0005$}
\def \eejjjjjjjOOqcd {$0.0157 \pm 0.0031$}
\def \eejjjjjjjOOtotal {$0.017 \pm 0.003$}
\def \eejjjjjjjOOdata {$0$}
\def \eemetjjOOdy {$3.7 \pm 0.8$}
\def \eemetjjOOqcd {$-$}
\def \eemetjjOOtotal {$3.7 \pm 0.8$}
\def \eemetjjOOdata {$2$}
\def \eemetjjjOOdy {$0.45 \pm 0.13$}
\def \eemetjjjOOqcd {$-$}
\def \eemetjjjOOtotal {$0.45 \pm 0.13$}
\def \eemetjjjOOdata {$1$}
\def \eemetjjjjOOdy {$0.061 \pm 0.028$}
\def \eemetjjjjOOqcd {$-$}
\def \eemetjjjjOOtotal {$0.061 \pm 0.028$}
\def \eemetjjjjOOdata {$1$}
\def \eemetjjjjjOOdy {$0.040 \pm 0.020$}
\def \eemetjjjjjOOqcd {$-$}
\def \eemetjjjjjOOtotal {$0.040 \pm 0.020$}
\def \eemetjjjjjOOdata {$0$}
\def \eemetjjjjjjOOdy {$0.008 \pm 0.004$}
\def \eemetjjjjjjOOqcd {$-$}
\def \eemetjjjjjjOOtotal {$0.008 \pm 0.004$}
\def \eemetjjjjjjOOdata {$0$}
\def \eemetjjjjjjjOOdy {$0.0010 \pm 0.0005$}
\def \eemetjjjjjjjOOqcd {$-$}
\def \eemetjjjjjjjOOtotal {$0.0010 \pm 0.0005$}
\def \eemetjjjjjjjOOdata {$0$}
\def \ZjjOOdy {$94 \pm 19$}
\def \ZjjOOqcd {$1.88 \pm 0.28$}
\def \ZjjOOtotal {$96 \pm 19$}
\def \ZjjOOdata {$82$}
\def \ZjjjOOdy {$12.7 \pm 2.7$}
\def \ZjjjOOqcd {$0.27 \pm 0.04$}
\def \ZjjjOOtotal {$13.0 \pm 2.7$}
\def \ZjjjOOdata {$11$}
\def \ZjjjjOOdy {$1.8 \pm 0.5$}
\def \ZjjjjOOqcd {$0.034 \pm 0.006$}
\def \ZjjjjOOtotal {$1.8 \pm 0.5$}
\def \ZjjjjOOdata {$1$}
\def \ZjjjjjOOdy {$0.26 \pm 0.10$}
\def \ZjjjjjOOqcd {$0.0025 \pm 0.0009$}
\def \ZjjjjjOOtotal {$0.26 \pm 0.10$}
\def \ZjjjjjOOdata {$0$}
\def \ZjjjjjjOOdy {$0.020 \pm 0.010$}
\def \ZjjjjjjOOqcd {$0.0036 \pm 0.0011$}
\def \ZjjjjjjOOtotal {$0.024 \pm 0.010$}
\def \ZjjjjjjOOdata {$0$}
\def \ZjjjjjjjOOdy {$0.0040 \pm 0.0020$}
\def \ZjjjjjjjOOqcd {$0.0019 \pm 0.0007$}
\def \ZjjjjjjjOOtotal {$0.0059 \pm 0.0021$}
\def \ZjjjjjjjOOdata {$0$}

Refs.~\cite{SherlockPRD1,SherlockPRD2} provide examples of \Sleuth's performance on representative signatures.  When ignorance of both $WW$ and $t\bar{t}$ is feigned in the $e\mu X$ final states, we find $\scriptP_{e\mu\met}=2.4\sigma$ and $\scriptP_{e\mu\met \jj}=2.3\sigma$ in D\O\ data, correctly indicating the presence of $WW$ and $t\bar{t}$.  When ignorance of $t\bar{t}$ only is feigned, we find $\scriptP_{e\mu\met \jj} = 1.9\sigma$.  Excesses are observed with only 3.9 $WW$ events expected in $e\mu\met$ (with a background of 45.6 events), and only 1.8 $t\bar{t}$ events in $e\mu\met \jj$ (with a background of 3.4 events), even though \Sleuth\ ``knows'' nothing about either $WW$ or $t\bar{t}$.  We are able to consistently find indications of the presence of $WW$ and $t\bar{t}$ in an ensemble of mock experiments at a similar level of sensitivity.  

In the $W$+jets-like final states we again feign ignorance of $t\bar{t}$ in the background estimate, and find $\scriptP_{\rm min}>3\sigma$ in 30\% of an ensemble of mock experimental runs on the final states $W\jjj$, $W\jjjj$, $W\jjjjj$, and $W\jjjjjj$.  In the $Z$+jets-like final states we consider a hypothetical signal: a first generation scalar leptoquark with a mass of 170~GeV and a branching ratio into charged leptons of $\beta=1$.  In the $ee\jj$ final state $5.9\pm0.8$ such leptoquark events would be expected with a background of \eejjOOtotal\ events.  \Sleuth\ finds $\scriptP_{ee\jj}>3.5\sigma$ in 80\% of the mock experiments performed.  Finally, in the final states $3\ellgamma X$ we find that a careful and systematic definition of final states can result in discovery sensitivity with only a few events, independent of their kinematics.  We conclude from these studies that \Sleuth\ is sensitive to a variety of new physics signatures.


Figure~\ref{fig:prl_data_plot_2} shows the results of the \Sleuth\ analysis of two typical final states ($W\jj$ and $Z\jj$).  The variable space defined by Table~\ref{tbl:VariableRules} is two-dimensional; parentheses are used in the axis labels to indicate the transformed variables of the unit box.  The circles are individual data events, and filled circles define the region selected by \Sleuth.  The regions chosen are seen to correspond to high $p_T$ in at least one dimension, as required by the imposed criteria.  Visually, these regions do not appear to contain an unusual excess, and large $\scriptP$s are found.  Similar results are obtained for other final states.

Table~\ref{tbl:FinalResults} summarizes the values of $\scriptP$ obtained for all populated final states analyzed in this article.  Taking into account the many final states (both populated and unpopulated) that are considered, we find $\gothicP$=\twiddleScriptPvalue, implying that 89\% of an ensemble of hypothetical similar experimental runs would have produced a final state with a candidate signal more interesting than the most interesting observed in these data.  Figure~\ref{fig:results_plot_prl}(b) shows a histogram of the $\scriptP$ values, in units of standard deviations, computed for the populated final states analyzed in this article, together with the distribution expected from a simulation of many mock experimental runs.  Good agreement is observed.  We find no evidence of new high $p_T$ physics in these data.

 \def \EMMDA {39}
 \def \EMMFA {18.4$\pm$1.4}
 \def \EMMTT {0.011$\pm$0.003}
 \def \EMMDY {0.5$\pm$0.2}
 \def \EMMWW {3.9$\pm$1.0}
 \def \EMMZT {25.6$\pm$6.5}
 \def \EMMTOT {48.5$\pm$7.6}
 \def \EMMJDA {13}
 \def \EMMJFA {8.7$\pm$1.0}
 \def \EMMJTT {0.4$\pm$0.1}
 \def \EMMJDY {0.1$\pm$0.03}
 \def \EMMJWW {1.1$\pm$0.3}
 \def \EMMJZT {3$\pm$0.8}
 \def \EMMJTOT {13.2$\pm$1.5}
 \def \EMMJJDA {5}
 \def \EMMJJFA {2.7$\pm$0.6}
 \def \EMMJJTT {1.8$\pm$0.5}
 \def \EMMJJDY {0.012$\pm$0.006}
 \def \EMMJJWW {0.18$\pm$0.05}
 \def \EMMJJZT {0.5$\pm$0.2}
 \def \EMMJJTOT {5.2$\pm$0.8}
 \def \EMMJJJDA {1}
 \def \EMMJJJFA {0.4$\pm$0.2}
 \def \EMMJJJTT {0.7$\pm$0.2}
 \def \EMMJJJDY {0.005$\pm$0.004}
 \def \EMMJJJWW {0.032$\pm$0.009}
 \def \EMMJJJZT {0.07$\pm$0.05}
 \def \EMMJJJTOT {1.3$\pm$0.3}
 \def \EMMJJJJDA {0}
 \def \EMMJJJJTT {0.16$\pm$0.04}
 \def \EMMJJJJDY {0.002$\pm$0.003}
 \def \EMMJJJJWW {0.004$\pm$0.002}
 \def \EMMJJJJZT {0.02$\pm$0.03}
 \def \EMMJJJJTOT {0.2$\pm$0.2}
 \def \EMMJJJJJDA {0}
 \def \EMMJJJJJFA {0$\pm$0.2}
 \def \EMMJJJJJTT {0.025$\pm$0.007}
 \def \EMMJJJJJDY {0$\pm$0.003}
 \def \EMMJJJJJWW {0$\pm$0.0006}
 \def \EMMJJJJJZT {0$\pm$0.03}
 \def \EMMJJJJJTOT {0.025$\pm$0.2}
 \def \EMXDA {58}
 \def \EMXFA {30.2$\pm$1.8}
 \def \EMXTT {3.1$\pm$0.5}
 \def \EMXDY {0.7$\pm$0.1}
 \def \EMXWW {5.2$\pm$0.8}
 \def \EMXZT {29.2$\pm$4.5}
 \def \EMXTOT {68.3$\pm$5.7}

\def \emumetOOtotal {$0.00 \pm 0.00$}
\def \emumetOOdata {$39$}
\def \emumetjOOtotal {$0.00 \pm 0.00$}
\def \emumetjOOdata {$13$}
\def \emumetjjOOtotal {$0.00 \pm 0.00$}
\def \emumetjjOOdata {$5$}
\def \emumetjjjOOtotal {$0.00 \pm 0.00$}
\def \emumetjjjOOdata {$1$}
\def \emumetjjjjOOtotal {$0.00 \pm 0.00$}
\def \emumetjjjjOOdata {$0$}
\def \emumetjjjjjOOtotal {$0.00 \pm 0.00$}
\def \emumetjjjjjOOdata {$0$}
\def \emumetjjjjjjOOtotal {$0.00 \pm 0.00$}
\def \emumetjjjjjjOOdata {$0$}
\def \emumetjjjjjjjOOtotal {$0.00 \pm 0.00$}
\def \emumetjjjjjjjOOdata {$0$}
\def \emumetgOOtotal {$0.00 \pm 0.00$}
\def \emumetgOOdata {$0$}
\def \emumetgjOOtotal {$0.00 \pm 0.00$}
\def \emumetgjOOdata {$0$}
\def \emumetgjjOOtotal {$0.00 \pm 0.00$}
\def \emumetgjjOOdata {$0$}
\def \emumetgjjjOOtotal {$0.00 \pm 0.00$}
\def \emumetgjjjOOdata {$0$}

\def \eeeOOzg {$2.6 \pm 1.0$}
\def \eeeOOegg {$-$}
\def \eeeOOggg {$-$}
\def \eeeOOzj {$-$}
\def \eeeOOwgg {$-$}
\def \eeeOOwz {$-$}
\def \eeeOOtotal {$2.6 \pm 1.0$}
\def \eeeOOdata {$1$}
\def \eemumetOOzg {$-$}
\def \eemumetOOegg {$-$}
\def \eemumetOOggg {$-$}
\def \eemumetOOzj {$-$}
\def \eemumetOOwgg {$-$}
\def \eemumetOOwz {$0.10 \pm 0.05$}
\def \eemumetOOtotal {$0.10 \pm 0.05$}
\def \eemumetOOdata {$0$}
\def \emumuOOzg {$-$}
\def \emumuOOegg {$-$}
\def \emumuOOggg {$-$}
\def \emumuOOzj {$-$}
\def \emumuOOwgg {$-$}
\def \emumuOOwz {$0.040 \pm 0.020$}
\def \emumuOOtotal {$0.040 \pm 0.020$}
\def \emumuOOdata {$0$}
\def \mumumuOOzg {$-$}
\def \mumumuOOegg {$-$}
\def \mumumuOOggg {$-$}
\def \mumumuOOzj {$-$}
\def \mumumuOOwgg {$-$}
\def \mumumuOOwz {$0.020 \pm 0.010$}
\def \mumumuOOtotal {$0.020 \pm 0.010$}
\def \mumumuOOdata {$0$}
\def \eegOOzg {$-$}
\def \eegOOegg {$-$}
\def \eegOOggg {$-$}
\def \eegOOzj {$-$}
\def \eegOOwgg {$-$}
\def \eegOOwz {$-$}
\def \eegOOtotal {$-$}
\def \eegOOdata {$1$}
\def \eegmetOOzg {$-$}
\def \eegmetOOegg {$-$}
\def \eegmetOOggg {$-$}
\def \eegmetOOzj {$-$}
\def \eegmetOOwgg {$-$}
\def \eegmetOOwz {$-$}
\def \eegmetOOtotal {$-$}
\def \eegmetOOdata {$1$}
\def \ZgOOzg {$-$}
\def \ZgOOegg {$-$}
\def \ZgOOggg {$-$}
\def \ZgOOzj {$-$}
\def \ZgOOwgg {$-$}
\def \ZgOOwz {$-$}
\def \ZgOOtotal {$-$}
\def \ZgOOdata {$3$}
\def \ZgjOOzg {$-$}
\def \ZgjOOegg {$-$}
\def \ZgjOOggg {$-$}
\def \ZgjOOzj {$-$}
\def \ZgjOOwgg {$-$}
\def \ZgjOOwz {$-$}
\def \ZgjOOtotal {$-$}
\def \ZgjOOdata {$1$}
\def \eggOOzg {$-$}
\def \eggOOegg {$10.7 \pm 2.1$}
\def \eggOOggg {$-$}
\def \eggOOzj {$-$}
\def \eggOOwgg {$-$}
\def \eggOOwz {$-$}
\def \eggOOtotal {$10.7 \pm 2.1$}
\def \eggOOdata {$6$}
\def \eggjOOzg {$2.3 \pm 0.7$}
\def \eggjOOegg {$-$}
\def \eggjOOggg {$-$}
\def \eggjOOzj {$-$}
\def \eggjOOwgg {$-$}
\def \eggjOOwz {$-$}
\def \eggjOOtotal {$2.3 \pm 0.7$}
\def \eggjOOdata {$4$}
\def \eggjjOOzg {$0.37 \pm 0.15$}
\def \eggjjOOegg {$-$}
\def \eggjjOOggg {$-$}
\def \eggjjOOzj {$-$}
\def \eggjjOOwgg {$-$}
\def \eggjjOOwz {$-$}
\def \eggjjOOtotal {$0.37 \pm 0.15$}
\def \eggjjOOdata {$1$}
\def \eggmetOOzg {$-$}
\def \eggmetOOegg {$-$}
\def \eggmetOOggg {$-$}
\def \eggmetOOzj {$-$}
\def \eggmetOOwgg {$0.026 \pm 0.010$}
\def \eggmetOOwz {$0.11 \pm 0.05$}
\def \eggmetOOtotal {$0.14 \pm 0.05$}
\def \eggmetOOdata {$1$}
\def \WggOOzg {$-$}
\def \WggOOegg {$-$}
\def \WggOOggg {$-$}
\def \WggOOzj {$-$}
\def \WggOOwgg {$0.045 \pm 0.015$}
\def \WggOOwz {$0.16 \pm 0.08$}
\def \WggOOtotal {$0.21 \pm 0.08$}
\def \WggOOdata {$1$}
\def \mumugOOzg {$3.9 \pm 0.9$}
\def \mumugOOegg {$-$}
\def \mumugOOggg {$-$}
\def \mumugOOzj {$-$}
\def \mumugOOwgg {$-$}
\def \mumugOOwz {$-$}
\def \mumugOOtotal {$3.9 \pm 0.9$}
\def \mumugOOdata {$0$}
\def \mumugjOOzg {$0.64 \pm 0.20$}
\def \mumugjOOegg {$-$}
\def \mumugjOOggg {$-$}
\def \mumugjOOzj {$-$}
\def \mumugjOOwgg {$-$}
\def \mumugjOOwz {$-$}
\def \mumugjOOtotal {$0.64 \pm 0.20$}
\def \mumugjOOdata {$0$}
\def \mumugjjOOzg {$0.13 \pm 0.04$}
\def \mumugjjOOegg {$-$}
\def \mumugjjOOggg {$-$}
\def \mumugjjOOzj {$-$}
\def \mumugjjOOwgg {$-$}
\def \mumugjjOOwz {$-$}
\def \mumugjjOOtotal {$0.13 \pm 0.04$}
\def \mumugjjOOdata {$0$}
\def \mumugjjjOOzg {$0.025 \pm 0.010$}
\def \mumugjjjOOegg {$-$}
\def \mumugjjjOOggg {$-$}
\def \mumugjjjOOzj {$-$}
\def \mumugjjjOOwgg {$-$}
\def \mumugjjjOOwz {$-$}
\def \mumugjjjOOtotal {$0.025 \pm 0.010$}
\def \mumugjjjOOdata {$0$}
\def \gggOOzg {$-$}
\def \gggOOegg {$-$}
\def \gggOOggg {$2.5 \pm 0.5$}
\def \gggOOzj {$-$}
\def \gggOOwgg {$-$}
\def \gggOOwz {$-$}
\def \gggOOtotal {$2.5 \pm 0.5$}
\def \gggOOdata {$2$}

\def \ZOOzg {$8.9 \pm 1.8$}
\def \ZOOzj {$0.021 \pm 0.006$}
\def \ZOOwz {$-$}
\def \ZOOtotal {$8.9 \pm 1.8$}
\def \ZOOdata {$5$}
\def \ZjOOzg {$2.1 \pm 0.6$}
\def \ZjOOzj {$0.0047 \pm 0.0014$}
\def \ZjOOwz {$-$}
\def \ZjOOtotal {$2.1 \pm 0.6$}
\def \ZjOOdata {$0$}
\def \ZjjOOzg {$-$}
\def \ZjjOOzj {$0.00054 \pm 0.00025$}
\def \ZjjOOwz {$-$}
\def \ZjjOOtotal {$0.00054 \pm 0.00025$}
\def \ZjjOOdata {$0$}
\def \ZgOOzg {$3.3 \pm 0.7$}
\def \ZgOOzj {$0.99 \pm 0.27$}
\def \ZgOOwz {$-$}
\def \ZgOOtotal {$4.3 \pm 0.7$}
\def \ZgOOdata {$3$}
\def \ZgjOOzg {$0.80 \pm 0.30$}
\def \ZgjOOzj {$0.23 \pm 0.06$}
\def \ZgjOOwz {$-$}
\def \ZgjOOtotal {$1.03 \pm 0.31$}
\def \ZgjOOdata {$1$}
\def \ZgjjOOzg {$0.10 \pm 0.05$}
\def \ZgjjOOzj {$0.033 \pm 0.009$}
\def \ZgjjOOwz {$-$}
\def \ZgjjOOtotal {$0.13 \pm 0.05$}
\def \ZgjjOOdata {$0$}
\def \ZgjjjOOzg {$0.020 \pm 0.010$}
\def \ZgjjjOOzj {$0.0048 \pm 0.0014$}
\def \ZgjjjOOwz {$-$}
\def \ZgjjjOOtotal {$0.025 \pm 0.010$}
\def \ZgjjjOOdata {$0$}
\def \ZgjjjjOOzg {$0.0040 \pm 0.0020$}
\def \ZgjjjjOOzj {$0.0009 \pm 0.0004$}
\def \ZgjjjjOOwz {$-$}
\def \ZgjjjjOOtotal {$0.0049 \pm 0.0020$}
\def \ZgjjjjOOdata {$0$}
\def \eegOOzg {$2.1 \pm 0.4$}
\def \eegOOzj {$0.13 \pm 0.04$}
\def \eegOOwz {$-$}
\def \eegOOtotal {$2.2 \pm 0.4$}
\def \eegOOdata {$1$}
\def \eegjOOzg {$0.50 \pm 0.25$}
\def \eegjOOzj {$0.033 \pm 0.009$}
\def \eegjOOwz {$-$}
\def \eegjOOtotal {$0.53 \pm 0.25$}
\def \eegjOOdata {$0$}
\def \eegjjOOzg {$0.10 \pm 0.05$}
\def \eegjjOOzj {$0.0046 \pm 0.0014$}
\def \eegjjOOwz {$-$}
\def \eegjjOOtotal {$0.10 \pm 0.05$}
\def \eegjjOOdata {$0$}
\def \eegjjjOOzg {$-$}
\def \eegjjjOOzj {$-$}
\def \eegjjjOOwz {$-$}
\def \eegjjjOOtotal {$-$}
\def \eegjjjOOdata {$0$}
\def \eegmetOOzg {$0.010 \pm 0.005$}
\def \eegmetOOzj {$0.024 \pm 0.007$}
\def \eegmetOOwz {$0.23 \pm 0.10$}
\def \eegmetOOtotal {$0.26 \pm 0.10$}
\def \eegmetOOdata {$1$}
\def \eegjmetOOzg {$0.0020 \pm 0.0010$}
\def \eegjmetOOzj {$0.012 \pm 0.003$}
\def \eegjmetOOwz {$0.045 \pm 0.020$}
\def \eegjmetOOtotal {$0.059 \pm 0.020$}
\def \eegjmetOOdata {$0$}
\def \eegjjmetOOzg {$-$}
\def \eegjjmetOOzj {$0.0035 \pm 0.0011$}
\def \eegjjmetOOwz {$-$}
\def \eegjjmetOOtotal {$0.0035 \pm 0.0011$}
\def \eegjjmetOOdata {$0$}
\def \eegjjjmetOOzg {$-$}
\def \eegjjjmetOOzj {$0.0012 \pm 0.0005$}
\def \eegjjjmetOOwz {$-$}
\def \eegjjjmetOOtotal {$0.0012 \pm 0.0005$}
\def \eegjjjmetOOdata {$0$}

\def \emetjjOOwjj {$6.7 \pm 1.4$}
\def \emetjjOOqcd {$3.3 \pm 0.9$}
\def \emetjjOOtop {$1.7 \pm 0.6$}
\def \emetjjOOtotal {$11.6 \pm 1.7$}
\def \emetjjOOdata {$7$}
\def \emetjjjOOwjj {$1.0 \pm 0.4$}
\def \emetjjjOOqcd {$0.48 \pm 0.22$}
\def \emetjjjOOtop {$1.0 \pm 0.4$}
\def \emetjjjOOtotal {$2.5 \pm 0.6$}
\def \emetjjjOOdata {$5$}
\def \emetjjjjOOwjj {$0.15 \pm 0.11$}
\def \emetjjjjOOqcd {$0.38 \pm 0.19$}
\def \emetjjjjOOtop {$0.26 \pm 0.09$}
\def \emetjjjjOOtotal {$0.80 \pm 0.24$}
\def \emetjjjjOOdata {$2$}
\def \emetjjjjjOOwjj {$0.030 \pm 0.020$}
\def \emetjjjjjOOqcd {$0.08 \pm 0.04$}
\def \emetjjjjjOOtop {$0.042 \pm 0.017$}
\def \emetjjjjjOOtotal {$0.15 \pm 0.05$}
\def \emetjjjjjOOdata {$0$}
\def \WjjOOwjj {$334 \pm 51$}
\def \WjjOOqcd {$12.0 \pm 2.6$}
\def \WjjOOtop {$4.0 \pm 1.4$}
\def \WjjOOtotal {$350 \pm 51$}
\def \WjjOOdata {$387$}
\def \WjjjOOwjj {$57 \pm 9$}
\def \WjjjOOqcd {$3.4 \pm 0.9$}
\def \WjjjOOtop {$6.0 \pm 2.1$}
\def \WjjjOOtotal {$66 \pm 9$}
\def \WjjjOOdata {$56$}
\def \WjjjjOOwjj {$5.9 \pm 1.3$}
\def \WjjjjOOqcd {$1.1 \pm 0.4$}
\def \WjjjjOOtop {$3.9 \pm 1.4$}
\def \WjjjjOOtotal {$10.9 \pm 1.9$}
\def \WjjjjOOdata {$11$}
\def \WjjjjjOOwjj {$0.8 \pm 0.3$}
\def \WjjjjjOOqcd {$0.19 \pm 0.12$}
\def \WjjjjjOOtop {$0.73 \pm 0.26$}
\def \WjjjjjOOtotal {$1.8 \pm 0.4$}
\def \WjjjjjOOdata {$1$}
\def \WjjjjjjOOwjj {$0.12 \pm 0.06$}
\def \WjjjjjjOOqcd {$0.030 \pm 0.015$}
\def \WjjjjjjOOtop {$0.10 \pm 0.04$}
\def \WjjjjjjOOtotal {$0.25 \pm 0.07$}
\def \WjjjjjjOOdata {$1$}
\def \WjjjjjjjOOwjj {$0.020 \pm 0.010$}
\def \WjjjjjjjOOqcd {$0.0040 \pm 0.0020$}
\def \WjjjjjjjOOtop {$0.008 \pm 0.005$}
\def \WjjjjjjjOOtotal {$0.032 \pm 0.011$}
\def \WjjjjjjjOOdata {$0$}

\def \WjjOOwjjmc {$48 \pm 15$}
\def \WjjOOzjjmc {$1.6 \pm 0.4$}
\def \WjjOOwwmc {$0.5 \pm 0.3$}
\def \WjjOOtopmc {$0.42 \pm 0.14$}
\def \WjjOOtotal {$50 \pm 15$}
\def \WjjOOdata {$54$}
\def \WjjjOOwjjmc {$10 \pm 3$}
\def \WjjjOOzjjmc {$0.27 \pm 0.08$}
\def \WjjjOOwwmc {$0.41 \pm 0.26$}
\def \WjjjOOtopmc {$0.58 \pm 0.20$}
\def \WjjjOOtotal {$11 \pm 3$}
\def \WjjjOOdata {$11$}
\def \WjjjjOOwjjmc {$2.8 \pm 1.3$}
\def \WjjjjOOzjjmc {$0.022 \pm 0.011$}
\def \WjjjjOOwwmc {$-$}
\def \WjjjjOOtopmc {$0.61 \pm 0.21$}
\def \WjjjjOOtotal {$3.5 \pm 1.3$}
\def \WjjjjOOdata {$4$}

\def \WjjOOemetwjj {$334 \pm 51$}
\def \WjjOOemetqcd {$12.0 \pm 2.6$}
\def \WjjOOemettop {$4.0 \pm 1.4$}
\def \WjjOOmumetwjjmc {$48 \pm 15$}
\def \WjjOOmumetzjjmc {$1.6 \pm 0.4$}
\def \WjjOOmumetwwmc {$0.5 \pm 0.3$}
\def \WjjOOmumettopmc {$0.42 \pm 0.14$}
\def \WjjOOtotal {$400 \pm 53$}
\def \WjjOOdata {$441$}
\def \WjjjOOemetwjj {$57 \pm 9$}
\def \WjjjOOemetqcd {$3.4 \pm 0.9$}
\def \WjjjOOemettop {$6.0 \pm 2.1$}
\def \WjjjOOmumetwjjmc {$10 \pm 3$}
\def \WjjjOOmumetzjjmc {$0.27 \pm 0.08$}
\def \WjjjOOmumetwwmc {$0.41 \pm 0.26$}
\def \WjjjOOmumettopmc {$0.58 \pm 0.20$}
\def \WjjjOOtotal {$77 \pm 10$}
\def \WjjjOOdata {$67$}
\def \WjjjjOOemetwjj {$5.9 \pm 1.3$}
\def \WjjjjOOemetqcd {$1.1 \pm 0.4$}
\def \WjjjjOOemettop {$3.9 \pm 1.4$}
\def \WjjjjOOmumetwjjmc {$2.8 \pm 1.3$}
\def \WjjjjOOmumetzjjmc {$0.022 \pm 0.011$}
\def \WjjjjOOmumetwwmc {$-$}
\def \WjjjjOOmumettopmc {$0.61 \pm 0.21$}
\def \WjjjjOOtotal {$14.3 \pm 2.3$}
\def \WjjjjOOdata {$15$}
\def \WjjjjjOOemetwjj {$0.8 \pm 0.3$}
\def \WjjjjjOOemetqcd {$0.19 \pm 0.12$}
\def \WjjjjjOOemettop {$0.73 \pm 0.26$}
\def \WjjjjjOOmumetwjjmc {$-$}
\def \WjjjjjOOmumetzjjmc {$-$}
\def \WjjjjjOOmumetwwmc {$-$}
\def \WjjjjjOOmumettopmc {$-$}
\def \WjjjjjOOtotal {$1.8 \pm 0.4$}
\def \WjjjjjOOdata {$1$}
\def \WjjjjjjOOemetwjj {$0.12 \pm 0.06$}
\def \WjjjjjjOOemetqcd {$0.030 \pm 0.015$}
\def \WjjjjjjOOemettop {$0.10 \pm 0.04$}
\def \WjjjjjjOOmumetwjjmc {$-$}
\def \WjjjjjjOOmumetzjjmc {$-$}
\def \WjjjjjjOOmumetwwmc {$-$}
\def \WjjjjjjOOmumettopmc {$-$}
\def \WjjjjjjOOtotal {$0.25 \pm 0.07$}
\def \WjjjjjjOOdata {$1$}
\def \WjjjjjjjOOemetwjj {$0.020 \pm 0.010$}
\def \WjjjjjjjOOemetqcd {$0.0040 \pm 0.0020$}
\def \WjjjjjjjOOemettop {$0.008 \pm 0.005$}
\def \WjjjjjjjOOmumetwjjmc {$-$}
\def \WjjjjjjjOOmumetzjjmc {$-$}
\def \WjjjjjjjOOmumetwwmc {$-$}
\def \WjjjjjjjOOmumettopmc {$-$}
\def \WjjjjjjjOOtotal {$0.032 \pm 0.011$}
\def \WjjjjjjjOOdata {$0$}

\def \eejjOOdy {$20 \pm 4$}
\def \eejjOOqcd {$12.2 \pm 1.8$}
\def \eejjOOtotal {$32 \pm 4$}
\def \eejjOOdata {$32$}
\def \eejjjOOdy {$2.6 \pm 0.6$}
\def \eejjjOOqcd {$1.85 \pm 0.28$}
\def \eejjjOOtotal {$4.5 \pm 0.6$}
\def \eejjjOOdata {$4$}
\def \eejjjjOOdy {$0.40 \pm 0.20$}
\def \eejjjjOOqcd {$0.24 \pm 0.04$}
\def \eejjjjOOtotal {$0.64 \pm 0.20$}
\def \eejjjjOOdata {$3$}
\def \eejjjjjOOdy {$0.030 \pm 0.015$}
\def \eejjjjjOOqcd {$0.026 \pm 0.005$}
\def \eejjjjjOOtotal {$0.056 \pm 0.016$}
\def \eejjjjjOOdata {$0$}
\def \eejjjjjjOOdy {$0.0060 \pm 0.0030$}
\def \eejjjjjjOOqcd {$0.018 \pm 0.003$}
\def \eejjjjjjOOtotal {$0.024 \pm 0.005$}
\def \eejjjjjjOOdata {$0$}
\def \eejjjjjjjOOdy {$0.0010 \pm 0.0005$}
\def \eejjjjjjjOOqcd {$0.0157 \pm 0.0031$}
\def \eejjjjjjjOOtotal {$0.017 \pm 0.003$}
\def \eejjjjjjjOOdata {$0$}
\def \eemetjjOOdy {$3.7 \pm 0.8$}
\def \eemetjjOOqcd {$-$}
\def \eemetjjOOtotal {$3.7 \pm 0.8$}
\def \eemetjjOOdata {$2$}
\def \eemetjjjOOdy {$0.45 \pm 0.13$}
\def \eemetjjjOOqcd {$-$}
\def \eemetjjjOOtotal {$0.45 \pm 0.13$}
\def \eemetjjjOOdata {$1$}
\def \eemetjjjjOOdy {$0.061 \pm 0.028$}
\def \eemetjjjjOOqcd {$-$}
\def \eemetjjjjOOtotal {$0.061 \pm 0.028$}
\def \eemetjjjjOOdata {$1$}
\def \eemetjjjjjOOdy {$0.040 \pm 0.020$}
\def \eemetjjjjjOOqcd {$-$}
\def \eemetjjjjjOOtotal {$0.040 \pm 0.020$}
\def \eemetjjjjjOOdata {$0$}
\def \eemetjjjjjjOOdy {$0.008 \pm 0.004$}
\def \eemetjjjjjjOOqcd {$-$}
\def \eemetjjjjjjOOtotal {$0.008 \pm 0.004$}
\def \eemetjjjjjjOOdata {$0$}
\def \eemetjjjjjjjOOdy {$0.0010 \pm 0.0005$}
\def \eemetjjjjjjjOOqcd {$-$}
\def \eemetjjjjjjjOOtotal {$0.0010 \pm 0.0005$}
\def \eemetjjjjjjjOOdata {$0$}
\def \ZjjOOdy {$94 \pm 19$}
\def \ZjjOOqcd {$1.88 \pm 0.28$}
\def \ZjjOOtotal {$96 \pm 19$}
\def \ZjjOOdata {$82$}
\def \ZjjjOOdy {$12.7 \pm 2.7$}
\def \ZjjjOOqcd {$0.27 \pm 0.04$}
\def \ZjjjOOtotal {$13.0 \pm 2.7$}
\def \ZjjjOOdata {$11$}
\def \ZjjjjOOdy {$1.8 \pm 0.5$}
\def \ZjjjjOOqcd {$0.034 \pm 0.006$}
\def \ZjjjjOOtotal {$1.8 \pm 0.5$}
\def \ZjjjjOOdata {$1$}
\def \ZjjjjjOOdy {$0.26 \pm 0.10$}
\def \ZjjjjjOOqcd {$0.0025 \pm 0.0009$}
\def \ZjjjjjOOtotal {$0.26 \pm 0.10$}
\def \ZjjjjjOOdata {$0$}
\def \ZjjjjjjOOdy {$0.020 \pm 0.010$}
\def \ZjjjjjjOOqcd {$0.0036 \pm 0.0011$}
\def \ZjjjjjjOOtotal {$0.024 \pm 0.010$}
\def \ZjjjjjjOOdata {$0$}
\def \ZjjjjjjjOOdy {$0.0040 \pm 0.0020$}
\def \ZjjjjjjjOOqcd {$0.0019 \pm 0.0007$}
\def \ZjjjjjjjOOtotal {$0.0059 \pm 0.0021$}
\def \ZjjjjjjjOOdata {$0$}

\def \ZjjOOwjjmctwo {$-$}
\def \ZjjOOzjjmctwo {$2.2 \pm 0.4$}
\def \ZjjOOwwmctwo {$-$}
\def \ZjjOOtopmctwo {$0.050 \pm 0.020$}
\def \ZjjOOtotal {$2.3 \pm 0.4$}
\def \ZjjOOdata {$3$}
\def \ZjjjOOwjjmctwo {$-$}
\def \ZjjjOOzjjmctwo {$0.24 \pm 0.05$}
\def \ZjjjOOwwmctwo {$-$}
\def \ZjjjOOtopmctwo {$0.018 \pm 0.009$}
\def \ZjjjOOtotal {$0.26 \pm 0.06$}
\def \ZjjjOOdata {$1$}
\def \ZjjjjOOwjjmctwo {$-$}
\def \ZjjjjOOzjjmctwo {$0.022 \pm 0.009$}
\def \ZjjjjOOwwmctwo {$-$}
\def \ZjjjjOOtopmctwo {$0.006 \pm 0.004$}
\def \ZjjjjOOtotal {$0.028 \pm 0.010$}
\def \ZjjjjOOdata {$0$}
\def \mumujjOOwjjmctwo {$-$}
\def \mumujjOOzjjmctwo {$0.112 \pm 0.029$}
\def \mumujjOOwwmctwo {$0.25 \pm 0.13$}
\def \mumujjOOtopmctwo {$0.14 \pm 0.05$}
\def \mumujjOOtotal {$0.50 \pm 0.15$}
\def \mumujjOOdata {$2$}
\def \mumujjjOOwjjmctwo {$-$}
\def \mumujjjOOzjjmctwo {$0.007 \pm 0.004$}
\def \mumujjjOOwwmctwo {$0.06 \pm 0.04$}
\def \mumujjjOOtopmctwo {$0.065 \pm 0.025$}
\def \mumujjjOOtotal {$0.13 \pm 0.05$}
\def \mumujjjOOdata {$0$}
\def \mumujjjjOOwjjmctwo {$-$}
\def \mumujjjjOOzjjmctwo {$0.0036 \pm 0.0027$}
\def \mumujjjjOOwwmctwo {$0.010 \pm 0.005$}
\def \mumujjjjOOtopmctwo {$0.012 \pm 0.007$}
\def \mumujjjjOOtotal {$0.025 \pm 0.009$}
\def \mumujjjjOOdata {$0$}

\def \ZjjOOeedy {$94 \pm 19$}
\def \ZjjOOeeqcd {$1.88 \pm 0.28$}
\def \ZjjOOmumuwjjmctwo {$-$}
\def \ZjjOOmumuzjjmctwo {$2.2 \pm 0.4$}
\def \ZjjOOmumuwwmctwo {$-$}
\def \ZjjOOmumutopmctwo {$0.050 \pm 0.020$}
\def \ZjjOOtotal {$98 \pm 19$}
\def \ZjjOOdata {$85$}
\def \ZjjjOOeedy {$12.7 \pm 2.7$}
\def \ZjjjOOeeqcd {$0.27 \pm 0.04$}
\def \ZjjjOOmumuwjjmctwo {$-$}
\def \ZjjjOOmumuzjjmctwo {$0.24 \pm 0.05$}
\def \ZjjjOOmumuwwmctwo {$-$}
\def \ZjjjOOmumutopmctwo {$0.018 \pm 0.009$}
\def \ZjjjOOtotal {$13.2 \pm 2.7$}
\def \ZjjjOOdata {$12$}
\def \ZjjjjOOeedy {$1.8 \pm 0.5$}
\def \ZjjjjOOeeqcd {$0.034 \pm 0.006$}
\def \ZjjjjOOmumuwjjmctwo {$-$}
\def \ZjjjjOOmumuzjjmctwo {$0.022 \pm 0.009$}
\def \ZjjjjOOmumuwwmctwo {$-$}
\def \ZjjjjOOmumutopmctwo {$0.006 \pm 0.004$}
\def \ZjjjjOOtotal {$1.9 \pm 0.5$}
\def \ZjjjjOOdata {$1$}
\def \ZjjjjjOOeedy {$0.26 \pm 0.10$}
\def \ZjjjjjOOeeqcd {$0.0025 \pm 0.0009$}
\def \ZjjjjjOOmumuwjjmctwo {$-$}
\def \ZjjjjjOOmumuzjjmctwo {$-$}
\def \ZjjjjjOOmumuwwmctwo {$-$}
\def \ZjjjjjOOmumutopmctwo {$-$}
\def \ZjjjjjOOtotal {$0.26 \pm 0.10$}
\def \ZjjjjjOOdata {$0$}
\def \ZjjjjjjOOeedy {$0.020 \pm 0.010$}
\def \ZjjjjjjOOeeqcd {$0.0036 \pm 0.0011$}
\def \ZjjjjjjOOmumuwjjmctwo {$-$}
\def \ZjjjjjjOOmumuzjjmctwo {$-$}
\def \ZjjjjjjOOmumuwwmctwo {$-$}
\def \ZjjjjjjOOmumutopmctwo {$-$}
\def \ZjjjjjjOOtotal {$0.024 \pm 0.010$}
\def \ZjjjjjjOOdata {$0$}
\def \ZjjjjjjjOOeedy {$0.0040 \pm 0.0020$}
\def \ZjjjjjjjOOeeqcd {$0.0019 \pm 0.0007$}
\def \ZjjjjjjjOOmumuwjjmctwo {$-$}
\def \ZjjjjjjjOOmumuzjjmctwo {$-$}
\def \ZjjjjjjjOOmumuwwmctwo {$-$}
\def \ZjjjjjjjOOmumutopmctwo {$-$}
\def \ZjjjjjjjOOtotal {$0.0059 \pm 0.0021$}
\def \ZjjjjjjjOOdata {$0$}

\begin{table}[htb]
\centering
\begin{tabular}{lccl}
Final State  	& Bkg		& Data		& \multicolumn{1}{c}{$\scriptP$} \\ \hline
~\\
\multicolumn{4}{c}{$e\mu X$} \\
$e\mu\met$   	&\EMMTOT	&\EMMDA		& \scriptPemumet\	 (\scriptPemumetsigma$\sigma$) \\
$e\mu\met j$  	&\EMMJTOT	&\EMMJDA	& \scriptPemumetj\ 	 (\scriptPemumetjsigma$\sigma$) \\
$e\mu\met \jj$ 	&\EMMJJTOT	&\EMMJJDA	& \scriptPemumetjj\  (\scriptPemumetjjsigma$\sigma$) \\
$e\mu\met \jjj$  &\EMMJJJTOT	&\EMMJJJDA	& \scriptPemumetjjj\ (\scriptPemumetjjjsigma$\sigma$) \\
~\\
\multicolumn{4}{c}{$W$+jets-like} \\
$W \jj$		&\WjjOOtotal	&\WjjOOdata	& \scriptPWjj\	 (\scriptPWjjsigma$\sigma$)	\\
$W \jjj$	&\WjjjOOtotal	&\WjjjOOdata 	& \scriptPWjjj\	 (\scriptPWjjjsigma$\sigma$)	\\
$W \jjjj$	&\WjjjjOOtotal	&\WjjjjOOdata 	& \scriptPWjjjj\	 (\scriptPWjjjjsigma$\sigma$)	\\
$W \jjjjj$	&\WjjjjjOOtotal &\WjjjjjOOdata	& \scriptPWjjjjj\  (\scriptPWjjjjjsigma$\sigma$) \\
$W \jjjjjj$	&\WjjjjjjOOtotal&\WjjjjjjOOdata	& \scriptPWjjjjjj\  (\scriptPWjjjjjjsigma$\sigma$) \\
$e\met \jj$ 	&\emetjjOOtotal	&\emetjjOOdata	& \scriptPemetjj\  (\scriptPemetjjsigma$\sigma$) \\
$e\met \jjj$  	&\emetjjjOOtotal&\emetjjjOOdata & \scriptPemetjjj\  (\scriptPemetjjjsigma$\sigma$) \\ 
$e\met \jjjj$  	&\emetjjjjOOtotal&\emetjjjjOOdata& \scriptPemetjjjj\  (\scriptPemetjjjjsigma$\sigma$) \\ 
~\\
\multicolumn{4}{c}{$Z$+jets-like} \\
$Z \jj$		&\ZjjOOtotal	&\ZjjOOdata	& \scriptPZjj\  (\scriptPZjjsigma$\sigma$) \\
$Z \jjj$		&\ZjjjOOtotal	&\ZjjjOOdata	& \scriptPZjjj\  (\scriptPZjjjsigma$\sigma$) \\
$Z \jjjj$	&\ZjjjjOOtotal	&\ZjjjjOOdata	& \scriptPZjjjj\  (\scriptPZjjjjsigma$\sigma$) \\
$ee \jj$ 	&\eejjOOtotal	&\eejjOOdata	& \scriptPeejj\	 (\scriptPeejjsigma$\sigma$)	\\
$ee \jjj$ 	&\eejjjOOtotal	&\eejjjOOdata	& \scriptPeejjj\	 (\scriptPeejjjsigma$\sigma$)	\\
$ee \jjjj$ 	&\eejjjjOOtotal	&\eejjjjOOdata	& \scriptPeejjjj\  (\scriptPeejjjjsigma$\sigma$) \\
$ee\met \jj$ 	&\eemetjjOOtotal&\eemetjjOOdata	& \scriptPeemetjj\  (\scriptPeemetjjsigma$\sigma$) \\
$ee\met \jjj$ 	&\eemetjjjOOtotal&\eemetjjjOOdata& \scriptPeemetjjj\  (\scriptPeemetjjjsigma$\sigma$) \\
$ee\met \jjjj$ 	&\eemetjjjjOOtotal&\eemetjjjjOOdata& \scriptPeemetjjjj\  (\scriptPeemetjjjjsigma$\sigma$) \\
$\mu\mu \jj$ 	&\mumujjOOtotal	&\mumujjOOdata	& \scriptPmumujj\  (\scriptPmumujjsigma$\sigma$) \\
~\\
\multicolumn{4}{c}{$3\ellgamma X$} \\
$eee$ 		&\eeeOOtotal	&\eeeOOdata	& \scriptPeee\  (\scriptPeeesigma$\sigma$) \\
$Z\gamma$ 	&\ZgOOtotal	&\ZgOOdata	& \scriptPZg\  (\scriptPZgsigma$\sigma$) \\
$Z\gamma j$ 	&\ZgjOOtotal	&\ZgjOOdata	& \scriptPZgj\  (\scriptPZgjsigma$\sigma$) \\
$ee\gamma$ 	&\eegOOtotal	&\eegOOdata	& \scriptPeeg\  (\scriptPeegsigma$\sigma$) \\
$ee\gamma\met$	&\eegmetOOtotal	&\eegmetOOdata	& \scriptPeegmet\  (\scriptPeegmetsigma$\sigma$) \\
$e\gamma\gamma$ &\eggOOtotal	&\eggOOdata	& \scriptPegg\  (\scriptPeggsigma$\sigma$) \\
$e\gamma\gamma j$ &\eggjOOtotal	&\eggjOOdata	& \scriptPeggj\  (\scriptPeggjsigma$\sigma$) \\
$e\gamma\gamma \jj$ &\eggjjOOtotal&\eggjjOOdata	& \scriptPeggjj\  (\scriptPeggjjsigma$\sigma$) \\
$W\gamma\gamma$ &\WggOOtotal	&\WggOOdata	& \scriptPWgg\  (\scriptPWggsigma$\sigma$) \\
$\gamma\gamma\gamma$ &\gggOOtotal&\gggOOdata	& \scriptPggg\  (\scriptPgggsigma$\sigma$) \\
\hline
\raisebox{-.6ex}{$\gothicP$} &&& \twiddleScriptPvalue\ (\twiddleScriptPvalueSigma$\sigma$) \\ 
\end{tabular}
\caption{Summary of results.  The most interesting final state is found to be $ee\jjjj$, with $\scriptP=$ \scriptPeejjjj.  Upon taking into account the many final states we have considered in this analysis, we find $\gothicP=$ \twiddleScriptPvalue.}  
\label{tbl:FinalResults}
\end{table}

{\dofig{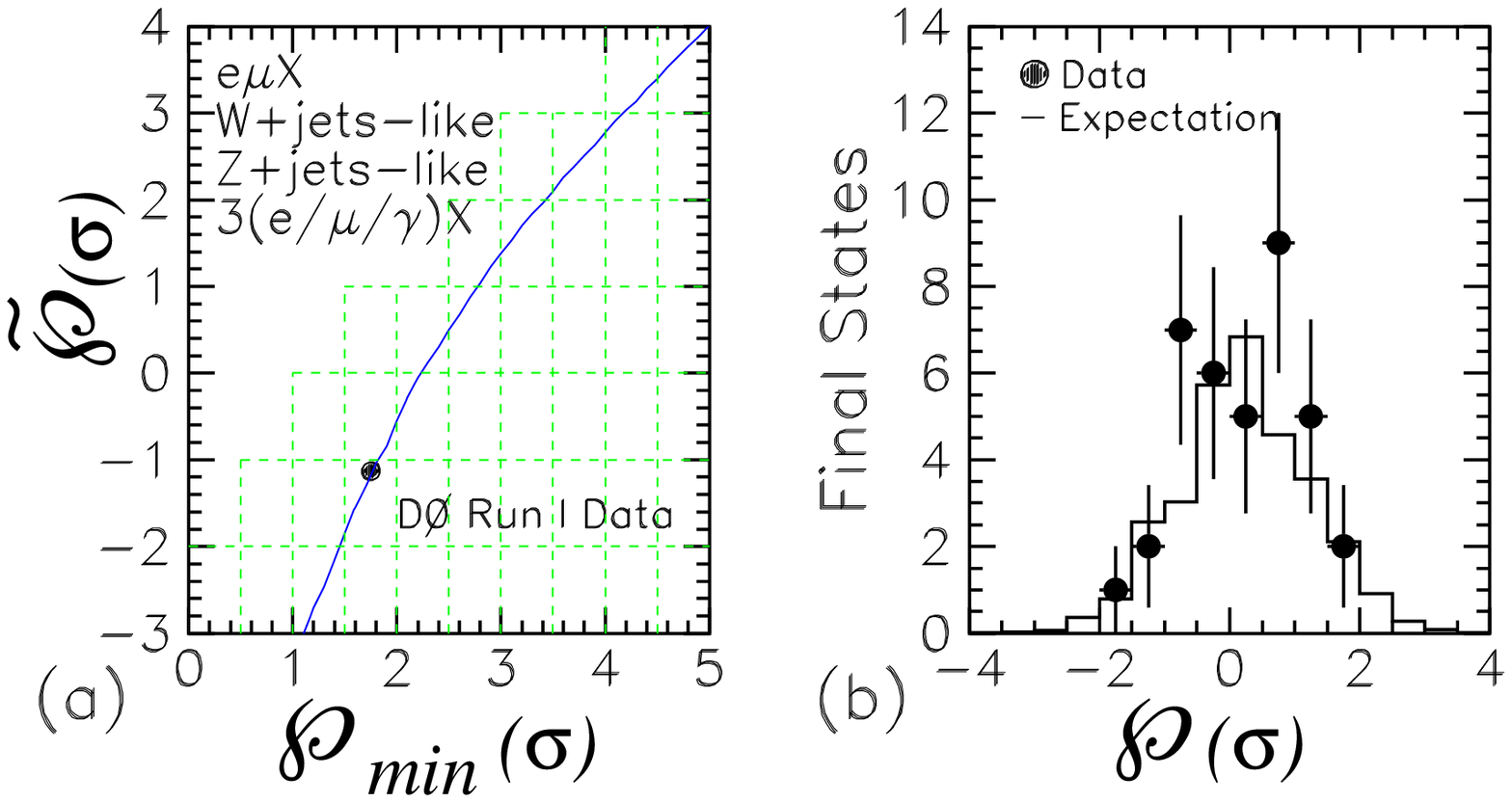} {3.5in}  {(a) The correspondence between $\gothicP$ and $\scriptP_{\rm min}$, each expressed in units of standard deviations.  The curve reflects the number of final states, both populated and unpopulated, considered in this Letter.  (b) Histogram of the $\scriptP$ values computed for the populated final states considered in this article, in units of standard deviations.  The distribution agrees well with expectation.} {fig:results_plot_prl} }

{\dofig{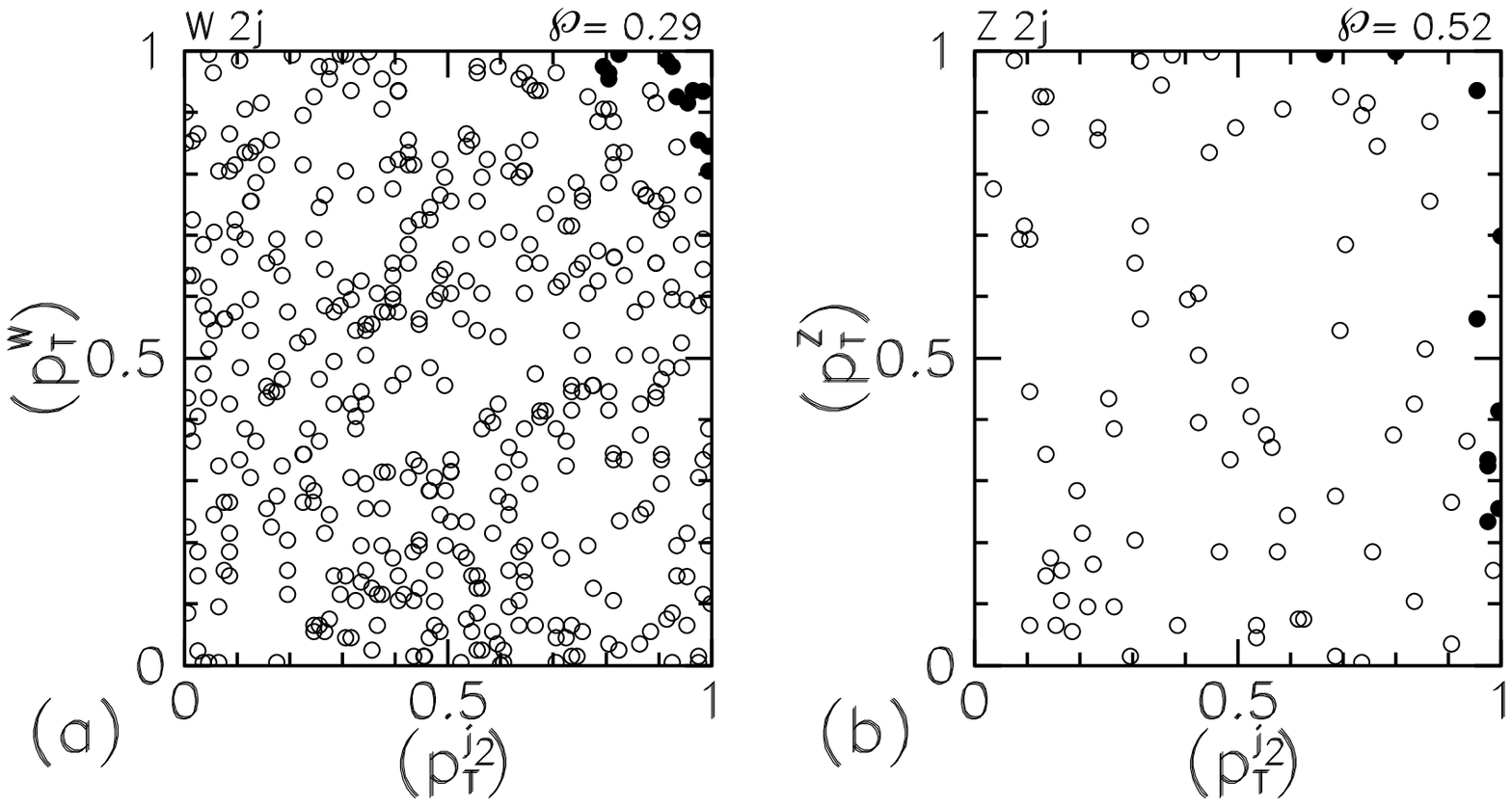} {3.5in} {Examples of \Sleuth's analysis of the final states (a) $W\jj$ and (b) $Z\jj$.} {fig:prl_data_plot_2} }


%
We thank the staffs at Fermilab and collaborating institutions, 
and acknowledge support from the 
Department of Energy and National Science Foundation (USA),  
Commissariat  \` a L'Energie Atomique and 
CNRS/Institut National de Physique Nucl\'eaire et 
de Physique des Particules (France), 
Ministry for Science and Technology and Ministry for Atomic 
   Energy (Russia),
CAPES and CNPq (Brazil),
Departments of Atomic Energy and Science and Education (India),
Colciencias (Colombia),
CONACyT (Mexico),
Ministry of Education and KOSEF (Korea),
CONICET and UBACyT (Argentina),
The Foundation for Fundamental Research on Matter (The Netherlands),
PPARC (United Kingdom),
A.P. Sloan Foundation,
and the A. von Humboldt Foundation.

\bibliographystyle{unsrt}

\begin{thebibliography}{2}

\bibitem{SherlockPRD1}
{D\O\ Collaboration},
B.~Abbott {\sl et al}.,
Phys. Rev. D
{\bf 62},
92004
(2000).

\bibitem{KnutesonThesis} 
B.~Knuteson, 
Ph.D. thesis,
University of California at Berkeley, 
2000 (unpublished).

\bibitem{SherlockPRD2}
{D\O\ Collaboration},
B.~Abbott {\sl et al}.,
submitted to Phys. Rev. D, 
hep-ex/0011067
(2000). 

\bibitem{D0Detector}
{D\O\ Collaboration},
S.~Abachi {\sl et al}.,
{Nucl. Instr. and Methods in Phys. Res. A}
{\bf 338},
{185} 
{(1994)}.

\bibitem{PreviousDZeroSearches}
{D\O\ Collaboration}, 
B.~Abbott {\sl et al}.,
Phys. Rev. Lett.
{\bf 82}, 
2457
(1999); 
{\sl ibid.} 
{\bf 80},
666 
(1998); 
{\sl ibid.} 
{\bf 82},
4769 
(1999); 
{D\O\ Collaboration}, 
S.~Abachi {\sl et al}.,
Phys. Lett. B
{\bf 385}, 
471 
(1996); 
{D\O\ Collaboration}, 
S.~Abachi {\sl et al}.,
Phys. Rev. Lett.
{\bf 76},
3271 
(1996); 
{\sl ibid.} 
{\bf 78},
3634
(1997); 
{D\O\ Collaboration}, 
B.~Abbott {\sl et al}.,
Phys. Rev. D
{\bf 61},
072001
(2000); 
{\sl ibid.} 
032004
(2000);
D\O\ Collaboration, 
B.~Abbott {\sl et al}.,
Phys. Rev. Lett.
{\bf 84},
2792
(2000);
{\sl ibid.} 
submitted to Phys. Rev. Lett.,
hep-ex/0010026
(2000). 

\bibitem{topCrossSection}
{D\O\ Collaboration},
S.~Abachi {\sl et al}.,
{Phys. Rev. Lett.} 
{\bf 79},
{1203}
{(1997)}.

\bibitem{Pseudorapidity}
The detector pseudorapidity $\eta_{\rm det}$ is defined with respect to the center of the detector, and the pseudorapidity $\eta$ with respect to the primary interaction point.

\bibitem{LeptoquarksToENu}
{D\O\ Collaboration},
B.~Abbott {\sl et al}.,
{Phys. Rev. Lett.}
{\bf 80},
2051
(1998).

\bibitem{LeptoquarksToMuNu}
{D\O\ Collaboration},
B.~Abbott {\sl et al}.,
{Phys. Rev. Lett.}
{\bf 83},
2896 
(1999).

\bibitem{LeptoquarksToEE}
{D\O\ Collaboration},
B.~Abbott {\sl et al}.,
{Phys. Rev. Lett.}
{\bf 79},
4321
(1997).

\bibitem{LeptoquarksToMuMu}
{D\O\ Collaboration},
B.~Abbott {\sl et al}.,
{Phys. Rev. Lett.}
{\bf 84},
2088 
(2000).

\end{thebibliography}

\end{document}